\documentstyle[aps,preprint,epsf]{revtex}
\begin{document}
\draft

\title{Far infrared spectroscopy on the three-dimensional dilute 
antiferromagnet Fe$_{x}$Zn$_{1-x}$F$_{2}$}
\author{J. Satooka and K. Katsumata}
\address{ RIKEN (The Institute of Physical and Chemical Research), 
Wako, Saitama 351-0198, Japan}
\author{D. P. Belanger}
\address{Department of Physics, University of California, Santa Cruz, 
California 95064}
\date{\today}           
\maketitle

\begin{abstract}
Fourier-transform Infrared (FT-IR) Spectroscopy measurements have been
performed on the three-dimensional dilute antiferromagnet
Fe$_{x}$Zn$_{1-x}$F$_{2}$ with $x = 0.99\sim 0.58$ in far infrared
(FIR) region.  The FIR spectra are analyzed taking into account
the ligand field and the local exchange interaction probability with
$J_{1}\sim J_{3}$; $|J_{1}|,|J_{3}|\ll|J_{2}|$, where $J_{1}$, 
$J_{2}$ and $J_{3}$ are the nearest neighbor, second nearest neighbor 
and third nearest neighbor exchange interaction constants, respectively.
The concentration dependence of the FIR spectra at low temperature is 
qualitatively well reproduced by our analysis, though some detailed structure remains
unexplained.  
\end{abstract}
\newpage

\section{Introduction} 
The dilute antiferromagnet Fe$_{x}$Zn$_{1-x}$F$_{2}$ has been studied 
extensively over the last two decades.
In the concentrated region with $x~>~0.4$, Fe$_{x}$Zn$_{1-x}$F$_{2}$ 
is an ideal example of the random-exchange Ising model without an 
applied field, and it becomes a prototypical example of the 
random-field Ising model when the field is applied along the spin 
easy axis\cite{Belanger}.
Near and below the percolation threshold $x_{p}\sim 
0.24$\cite{Sykes}, Fe$_{x}$Zn$_{1-x}$F$_{2}$ shows a cluster-glass 
behavior\cite{JS97,JS98}.
Paduani $et~al.$ have investigated the magnetic excitations in 
Fe$_{x}$Zn$_{1-x}$F$_{2}$ using inelastic neutron scattering 
techniques and have reported that the spectrum obtained at a low 
temperature can be reproduced with three Gaussian peaks\cite{Paduani}.
No theoretical explanation has been given for this.
In order to clarify the nature of the magnetic excitations in 
Fe$_{x}$Zn$_{1-x}$F$_{2}$, we have investigated the far infrared 
(FIR) spectra in Fe$_{x}$Zn$_{1-x}$F$_{2}$ using a Fourier-transform 
Infrared (FT-IR) spectrophotometer with high resolution in zero field 
and tried to understand the results.

\section{Experimental details}
The single crystals of Fe$_{x}$Zn$_{1-x}$F$_{2}$ were grown at UCSB 
by the Bridgman method.
The platelets were polished to form a 
wedge in order to avoid the interference between light reflected 
from the upper plane and that from the lower plane of the sample.
The strain at the surfaces of the sample, due to polishing, was 
removed by etching it in 70 $^{\circ }$C HCl solution for 1 minute.
The transition temperatures are $T_{N}$ = 78 K, 72 K, 66 K, 54 K, and 
45 K for $x$ = 0.99, 0.91, 0.84, 0.68, and 0.58, respectively.
The FIR spectra were measured with a bandwidth resolution of 0.1 
cm$^{-1}$ using a FT-IR spectrophotometer made by the Bruker 
company, Germany.
We used a high pressure Hg lamp as a light source, a myler film with 
23 $\mu m$ thickness as a beamsplitter, and a Si- bolometer as a 
detector.
The temperature, $T$,  of the sample was controlled using a helium cryostat 
from the Oxford Instruments, U. K.  

\section{Experimental results}
\begin{figure}
\epsfxsize 15cm
\centerline{
\epsfbox{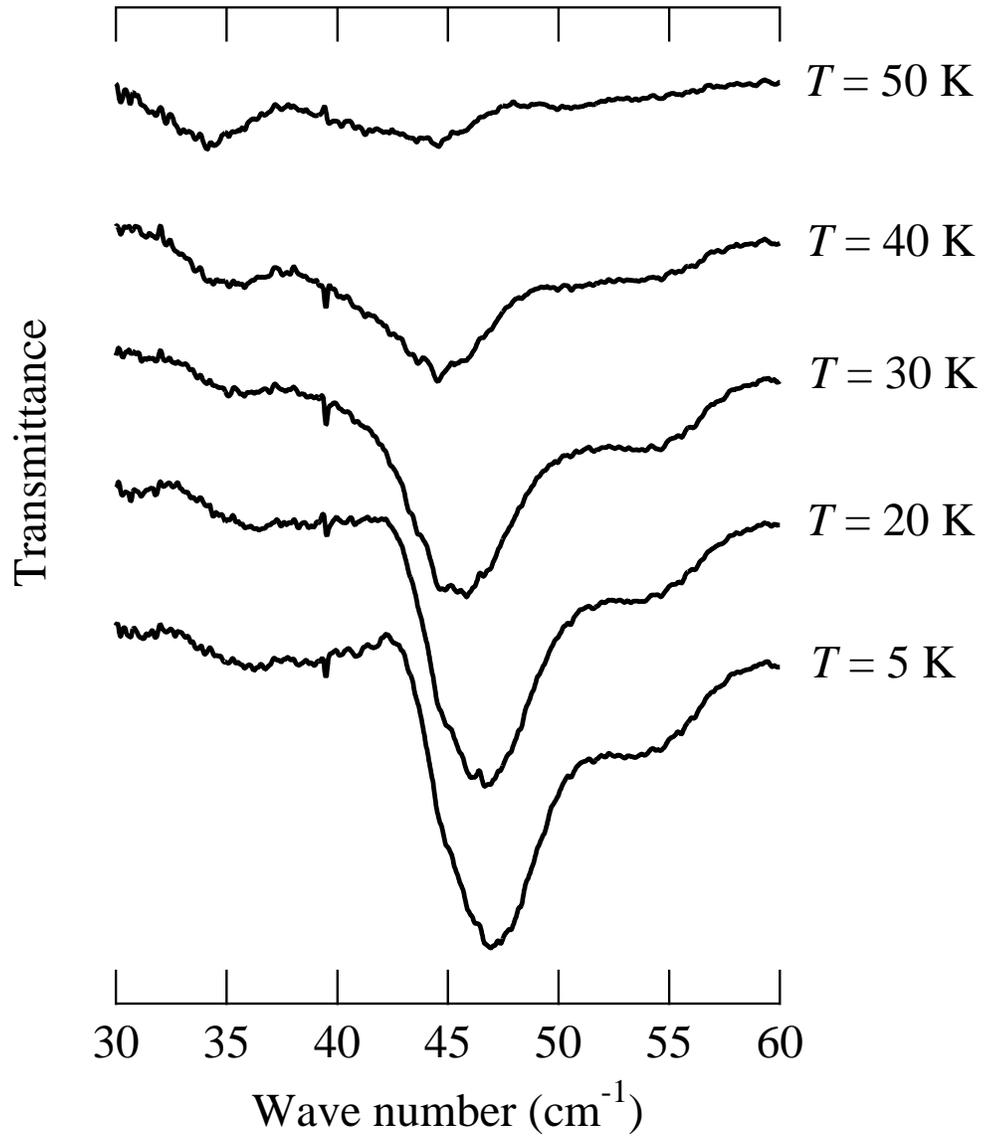}}
\caption{Temperature dependence of the FIR spectrum of
Fe$_{0.84}$Zn$_{0.16}$F$_{2}$.}
\label{temperature}
\end{figure}

The $T$ dependence of the FIR spectra of Fe$_{0.84}$Zn$_{0.16}$F$_{2}$
is shown in Fig. \ref{temperature} as an example.
The main broad absorption peak line becomes broader and its position
shifts gradually to the lower energy side as $T$ increases below $T_{N}$.  
Two smaller satellite peaks are visible at all temperatures. 
The higher energy one becomes broader as $T$ 
increases, while the lower energy one becomes sharper.  
Above $T_{N}$, no absorption is observed.
This $T$ dependence of the FIR spectra is found to be common
to all the samples investigated.

The concentration dependence of the FIR spectrum of
Fe$_{x}$Zn$_{1-x}$F$_{2}$ at 5 K is shown in Fig. \ref{concentration}.  
A very sharp absorption at
52.6 cm$^{-1}$ is observed for Fe$_{0.99}$Zn$_{0.01}$F$_{2}$.
The frequency of this absorption is in very good agreement with that 
of the antiferromagnetic resonance in FeF$_{2}$ (52.7 $\pm 0.2$ 
cm$^{-1}$) reported by Ohlman and Tinkham\cite{Ohlman}.  
However, there is a small absorption in the spectrum of Fe$_{0.99}$Zn$_{0.01}$F$_{2}$
near 50 cm$^{-1}$ which is not observed in the pure system.
The absorption line of Fe$_{0.91}$Zn$_{0.09}$F$_{2}$ is much
broader than that of Fe$_{0.99}$Zn$_{0.01}$F$_{2}$.
The oscillation in the spectrum of Fe$_{0.91}$Zn$_{0.09}$F$_{2}$
is considered to be an interference effect because this platelet is 
thicker than others.
With increasing dilution, the absorption line becomes broader and its
position shifts to lower energy.
Taking into account the higher resolution of the FIR data that allows
the main peak to be resolved into the two higher energy peaks, the spectrum of Fe$_{0.58}$Zn$_{0.42}$F$_{2}$ is consistent with 
the $q = 0$ spectrum of Fe$_{0.59}$Zn$_{0.41}$F$_{2}$ measured by 
neutron scattering\cite{Paduani}, except for the small 
lower energy peak in the FIR spectrum. 

\section{Discussion and conclusions}
To analyze the FIR spectra of Fe$_{x}$Zn$_{1-x}$F$_{2}$,
we approximate the magnetic excitations in this system as
single ion ones, taking the exchange interactions between Fe$^{2+}$ spins into 
account in the form of a molecular field.
This should be a good approximation since the results of the 
magnetization measurements made in Fe$_{x}$Zn$_{1-x}$F$_{2}$ were successfully
explained based on a localized spin flip model\cite{King81}.
The non-diluted compound FeF$_{2}$ has the rutile type crystal 
structure $D^{14}_{4h}-P4/mnm$\cite{Stout54,Stout55}.
The Fe$^{2+}$ free ion has a 3$d^{6}$ configuration and the ground 
state is $^{5}$D.
Each Fe$^{2+}$ ion is surrounded by six F$^{-}$ ions forming an 
octahedron.
The orbital state of the high spin Fe$^{2+}$ ion in an octahedral 
environment is split into the doublet $^{5}$E and the triplet $^{5}$T$_{2}$ by the
cubic field: $^{5}$T$_{2}$ is the ground state and $^{5}$E is lifted 
up by some 10,000 cm$^{-1}$.
The rhombic field removes all the orbital degeneracy of 
$^{5}$T$_{2}$: A$_{1g}$ is the ground state separated from B$_{1g}$ 
and B$_{2g}$ by $\sim $1115 cm$^{-1}$ and $\sim $2400 cm$^{-1}$ 
respectively\cite{Stout68}.
Spin-orbit effects are adequately treated by perturbation methods, 
the effective Hamiltonian pertaining to the lowest orbital state of a 
single Fe$^{2+}$ ion being
\begin{equation}
    {\cal H}=-D{S_{z}}^{2}+E({S_{x}}^{2}-{S_{y}}^{2}),
\label{eqn:Hamiltonian}
\end{equation}
where $z$ is taken parallel to the crystalline $c$ axis and $D$ and 
$E$ are the uniaxial and orthorhombic anisotropy constants.
As is seen from Fig. 1, the absorption intensity of the main signal 
decreases with increasing $T$.
This means that the signal originates in the transition from the 
ground state.
So, we assign the signals to
the transition, from the ground state to the lowest excitation state,
of which probability is suggested to be most dominant by matrix 
calculation of Hamiltonian of Eq. \ref{eqn:Hamiltonian}.
The excitation energy, $\Delta E$, for this transition to occur at
$T=0$ is 
\begin{equation}
    \mit\Delta E = \Delta E_{l}+|J|,
\label{eqn:energy}
\end{equation}
where $\Delta E_{l}=3(D-E)+3E^{2}/D$ is a ligand field and $J$ is an exchange 
interaction.
Tinkham estimated that $D=7.3 (\pm 0.7)$ cm$^{-1}$ and
$E=0.70 (\pm 0.04)$ cm$^{-1}$ from the analysis of the
paramagnetic resonance of Fe$^{2+}$ in ZnF$_{2}$\cite{Tim56}.
Guggenheim $et~al.$ estimated $D=6.46$ (+0.29, -0.10) cm$^{-1}$
by inelastic neutron scattering in FeF$_{2}$\cite{Gu68}.

\begin{figure}
\epsfxsize 15cm
\centerline{
\epsfbox{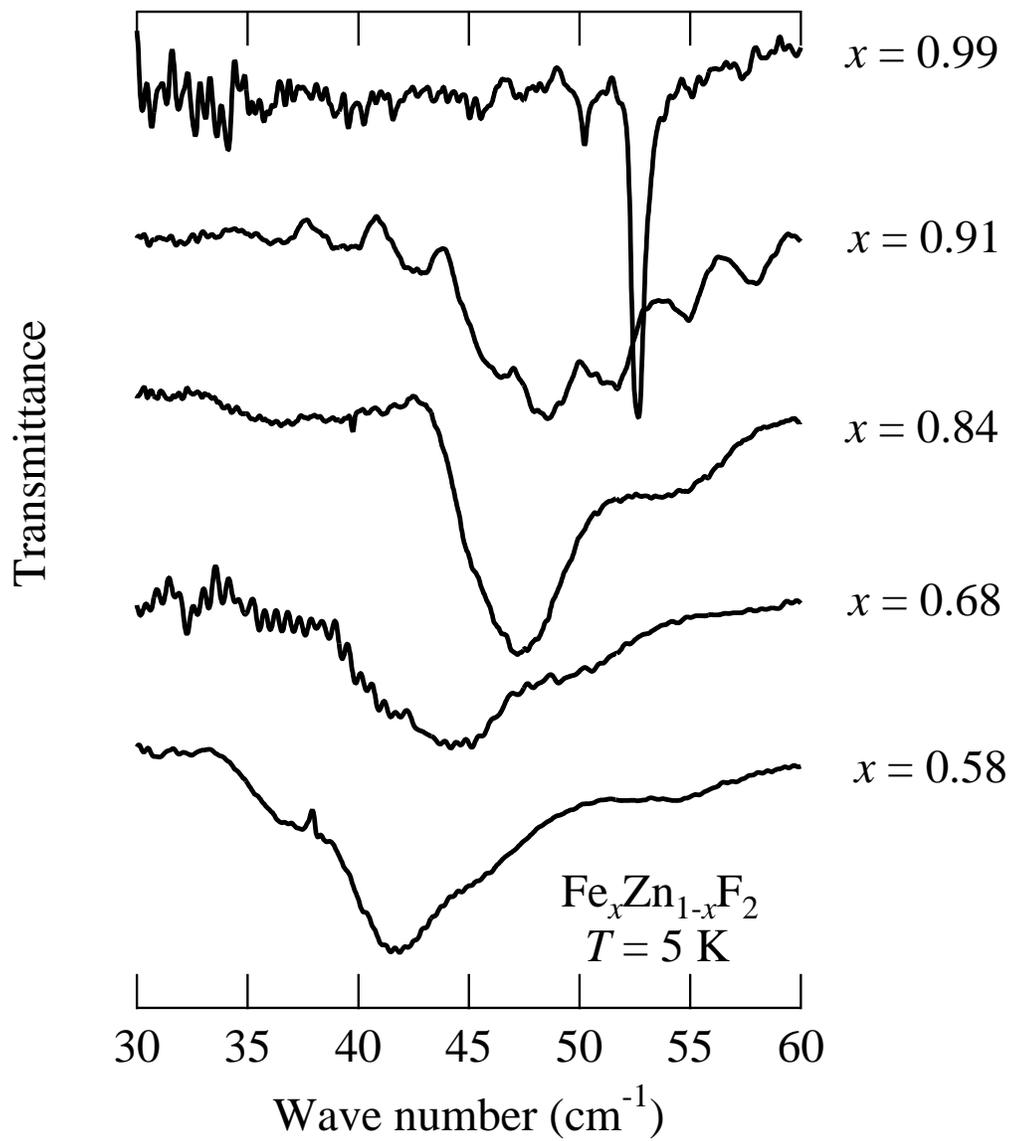}}
\caption{Concentration dependence of the FIR spectrum of
Fe$_{x}$Zn$_{1-x}$F$_{2}$ at 5 K. }
\label{concentration}
\end{figure}

\begin{figure}
\epsfxsize 15cm
\centerline{
\epsfbox{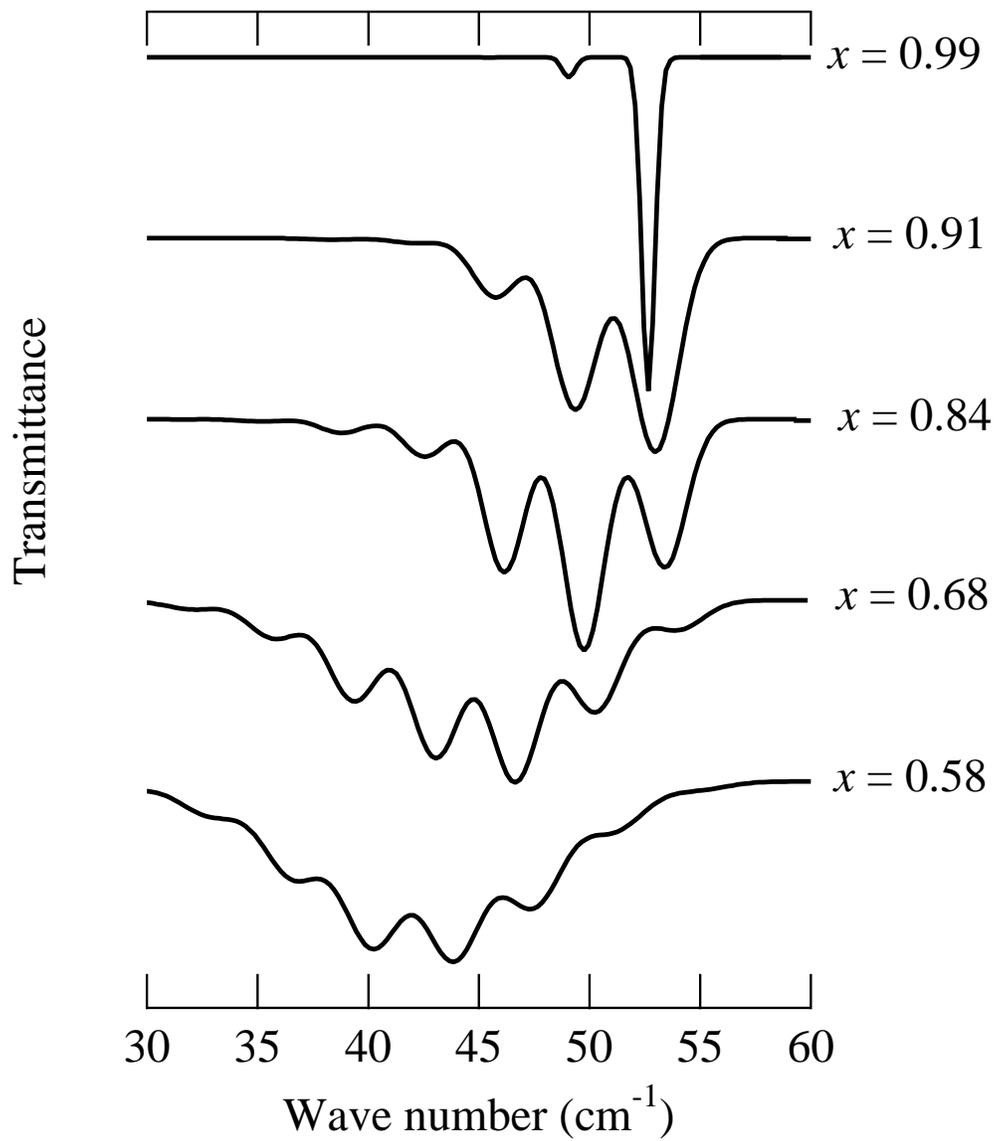}}
\caption{Concentration dependence of the simulated FIR spectrum
of Fe$_{x}$Zn$_{1-x}$F$_{2}$.}
\label{simulation}
\end{figure}

The probability, $p({i, n_{i}})$ that $n_{i}$ magnetic ions occupy the $z_{i}$ $i$ th
nearest neighbor sites is given by, 
$\frac{z_{i}!}{n_{i}!(z_{i}-n_{i})!}x^{n_{i}}(1-x)^{z_{i}-n_{i}}$,
with $i>0$ and where $x$ is the concentration of the magnetic ion.
The sum of the local exchange interaction,
$J({n_{i}}) = J({n_{1}, n_{2}, \cdots })$,
and the probability that $J({n_{i}})$ is operative,
$P({n_{i}}) = P({n_{1}, n_{2}, \cdots })$,
are described as $J({n_{i}}) = \sum_{i} J_{i}n_{i}$
and $P({ n_{i}}) = \prod_{i} p(i, n_{i})$,
where $J_{i}$ describes the exchange interaction to the $i$ th 
nearest neighbor.
In the calculation, we consider the exchange interactions up to the
third nearest neighbors.
The values of parameters used are as follows: $J_{1}$ = +0.048 
cm$^{-1}$,
$J_{2}$ = -3.64 cm$^{-1}$, $J_{3}$ = -0.194 cm$^{-1}$\cite{Hu70},
$z_{1}$ = 2, $z_{2}$ = 8, and $z_{3}$ = 4.
Taking into account the intrinsic line width of the experimental 
spectra, we calculate the spectrum by replacing $J$ in Eq. 
\ref{eqn:energy} by the Gaussian distribution of the probability of 
the local exchange interaction.
The absorption spectrum as a function of energy, $A(E)$, is written 
readily as
\begin{equation}
    A(E) = {\Large \int} \frac{P(|J|-\mit \Delta E_{l})}{\sqrt{2\pi 
}}exp\left\{\frac{E-(|J|-\mit \Delta E_{l})}{2\mit \sigma 
^{2}}\right\}d|J|,
\label{eqn:absorption}
\end{equation}
where $\Delta E_{l}$ and the width of the Gaussian distribution, 
$\sigma $, are parameters depending on the iron concentration $x$.
The concentration dependence of the FIR spectrum of 
Fe$_{x}$Zn$_{1-x}$F$_{2}$ calculated using Eq. \ref{eqn:absorption} is 
shown in Fig. \ref{simulation}.
These spectra are qualitatively in good agreement with the 
main peak in the experimental data shown in Fig. \ref{concentration}.
The molecular field approximation does not reproduce the three peaks
structure which was suggested from neutron scattering, 
but confirms the basics of the magnetic excitations of Fe$_{x}$Zn$_{1-x}$F$_{2}$.

In conclusion, we have studied the temperature dependence and 
concentration dependence of the FIR spectra of Fe$_{x}$Zn$_{1-x}$F$_{2}$.
The overall width and position of the main cluster of peaks in each of the low temperature FIR 
spectra is reproduced qualitatively by a calculation based on a single-ion
excitation model under a molecular field with the distribution of the
local exchange interactions.  The detailed structure is, however, not adequately
explained and further theoretical work is warranted.  In particular, the positions
of the resolved multiple FIR peaks are not reproduced by the model, consistent with conclusions of the previous neutron scattering study\cite{Paduani} at $x=0.59$.

\begin{acknowledgments}
We would like to thank A. Fukaya, T. Mutou and K. Hashi for 
helpful discussions.
This work was supported by the MR Science Research Program of RIKEN 
and DOE Grant No. DE-FG03-87ER45324.
One of the present authors (J. S.) is supported by the Special 
Postdoctoral Researchers Program from RIKEN. 
\end{acknowledgments}

\newpage

\newpage

\end{document}